\documentclass[sigconf]{acmart}

\usepackage{booktabs} 

\usepackage{graphicx,color,colortbl,psfrag,subfig}
\usepackage{multirow}
\usepackage{flushend}
\usepackage{algorithm}
\usepackage{algpseudocode}

\usepackage{amssymb}
\usepackage{amsmath}
\usepackage{amsfonts}

\usepackage{arydshln}

\begin{document}
\title[]{Towards Global Optimization in Display Advertising by Integrating Multimedia Metrics with Real-Time Bidding}

\author{Xiang Chen}
\affiliation{\institution{School of Computing\\National University of Singapore, Singapore, 117417}}
\email{chxiang@comp.nus.edu.sg}

\begin{abstract}

Real-time bidding (RTB) has become a new norm in display advertising where a publisher uses auction models to sell online user's page view to advertisers. In RTB, the ad with the highest bid price will be displayed to the user. This ad displaying process is biased towards the publisher. In fact, the benefits of the advertiser and the user have been rarely discussed. Towards the global optimization, we argue that all stakeholders' benefits should be considered. To this end, we propose a novel computation framework where multimedia techniques and auction theory are integrated. This doctoral research mainly focus on 1) figuring out the multimedia metrics that affect the effectiveness of online advertising; 2) integrating the discovered metrics into the RTB framework. We have presented some preliminary results and discussed the future directions.

\end{abstract}



\copyrightyear{2017}
\acmYear{2017}
\setcopyright{rightsretained}
\acmConference{MM'17}{}{October 23--27, 2017, Mountain View, CA, USA.}
\acmPrice{}
\acmDOI{https://doi.org/10.1145/3123266.3123966}
\acmISBN{ISBN 978-1-4503-4906-2/17/10}

\fancyhead{}

\maketitle

\section{Introduction}
\label{sec:introduction}
The explosive growth of multimedia data on Internet has created huge opportunities for online advertising. According to the latest Interactive Advertising Bureau's annual report~\cite{iab_2016}, the second-half revenue totaled \$39.8 billion in 2016, which is an increase of \$7.7 billion from the second-half revenue of 2015. Despite the increasing growth of revenue, the effectiveness of online advertising remains debatable. On the one hand, the user gets annoyed by the improper ads. Marketing data~\cite{hubspot_annoy_ad} shows that: the average click-through rate of display ads across all formats and placement is 0.06\%; young adults tend to ignore online banner ads; and ad blocking has grown by 41\% globally in the year 2015. The unpleasant ad experience further influences their site visits, and their perception of the displayed ad. On the other hand, the advertiser gets harmed by ineffective ad delivery. In terms of cost-per-impression pricing mechanism, the advertiser has to pay for every impression. Due to user's low engagement towards the displayed ad, a large amount of budget that the advertiser spends on online advertising is wasted.    
 
\begin{figure}[t]
\centering
\includegraphics[width=0.84\linewidth]{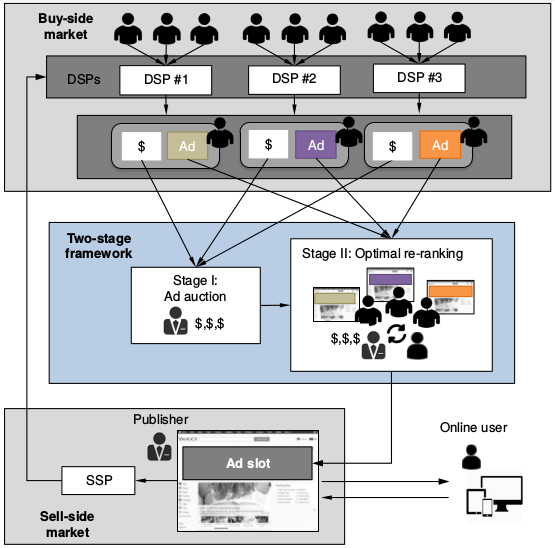}
\vspace{-7pt}
\caption{Overview of the proposed framework~\cite{chen2017RTBoptimizing}.}
\vspace{7pt}
\label{fig:schematic_view}
\end{figure}

The majority of online displayed ads are served through RTB. Compared to the RTB system that focuses on the publisher's revenue (local maximization), we define the global optimization as the optimal trade-offs among all stakeholders. In this regard, the proper ad is not only the highest revenue from the economic point of view, but also the best fit with the context from the multimedia point of view. Fig.~\ref{fig:schematic_view} illustrates how the multimedia metrics are incorporated with RTB. Our proposed framework consists of two stages: 1) the first stage uses the second-price (SP) auction model to deal with economic issues; 2) the second stage uses our proposed optimal re-ranking model to select the proper ad. Note that, the independent process of stage I ensures the equilibrium in ad auctions. The multimedia metrics take effect in stage II. Since they are exogenous, our proposed framework does not affect the advertiser's bidding behavior.

Due to the complex nature of display advertising, we are facing the following challenges:  
\begin{itemize}
\item \textbf{What metrics to select?} User's engagement (e.g., view, click) towards the displayed ad varies in different context. Understanding the context and user's behavior becomes the key factor to improve the effectiveness of online advertising.   
\item \textbf{How to integrate the metrics?} Online advertising is a multidisciplinary topic, which involves economics, multimedia and psychology. The metrics represent the stakeholders' benefit from different domains. An easy but effective solution is the linear combination model. Thus, determining the optimal weights becomes the key issue.
\end{itemize}

The development of computer vision, multimedia analysis and machine learning techniques brings opportunities to solve these two challenges. Firstly, we select a set of multimedia metrics according to marketing data and consumer psychology literature. We then conduct a user-study experiment to demonstrate the effectiveness of the selected metrics. Secondly, we simulate a real-world advertising system based on an auction dataset and two multimedia datasets. By adding constrains on the changes of metrics when compared with existing system, we propose an optimization model to obtain the optimal weights. To the best of our knowledge, we are the first towards global optimization in display advertising by combining multimedia techniques and auction theory. 

The rest of the paper is organized as follows. Section~\ref{sec:related_work} reviews the related work. Section~\ref{sec:approach} describes the proposed optimization framework. Section~\ref{sec:experiments}  presents our preliminary results. Section~\ref{sec:work_in_progress} lists our works in progress. And section~\ref{sec:conclusion} concludes the paper.

\section{Related work}
\label{sec:related_work}
From the economics perspective, researchers mainly focus on investigating the strategy of selling ad impressions. The generalized second-price (GSP) auction~\cite{Edelman_2007_2} and the Vickrey-Clark-Groves (VCG) auction~\cite{Parkes_2007} have been widely used on different platforms. As revenue is always the primary concern, the squashing parameter and reserve price have been discussed to increase the revenue of GSP auctions~\cite{Lahaie_2007, Thompson_2013}. GSP auction model has been proved to be able to improve social welfare as well~\cite{Lahaie_2011}. 

From the multimedia perspective, researchers mainly focus on increasing user's engagement towards displayed ads and meanwhile maintaining user's online experience. Literature in marketing and consumer psychology has already shown that the contextual relevance between the content of the hosting webpage and the ads makes a large difference in their clickability~\cite{chatterjee2003modeling}, and it also has a leading effect on user's online experience~\cite{mccoy2007effects}. According to the utilized information for contextual similarity matching, we can classify most existing contextual advertising systems into three categories: text-based advertising~\cite{li2010pagesense}, visual-based advertising~\cite{sengamedu2007vadeo} and targeted advertising~\cite{wang2009argo}. To achieve better contextual relevance, the multimodal approach has been proposed which considers both textual and visual information for video~\cite{mei2007videosense, guo2009adon} and image advertising~\cite{mei2012imagesense}. 

The works discussed so far only focus on the benefit of a specific stakeholder, regardless of the other stakeholders' benefits. Trade-offs among multiple objectives or stakeholders have been discussed in several works. Likhodedov et al.~\cite{Likhodedov_2003} proposed a framework that linearly combines revenue and social welfare in a single-item auction. Bachrach et al.~\cite{Bachrach_2014} proposed a framework that linearly combines relevant, welfare and clicks in sponsored search. Liao et al.~\cite{liao2008adimage} combined revenue and relevance for video advertising. 

Our goal is to foster a vigorous and healthy online advertising ecosystem. Different from most existing works that focus on local maximum value, we focus on the global optimization of the whole advertising ecosystem. Display advertising is an interplay among all stakeholders and each stakeholder's ad experience is an important factor for advertising effectiveness. In the proposed framework, the publisher's revenue will decrease in a short term in trade of improvements on the benefits of the advertiser and user. The increased ad experience will increase the demand of the advertiser's advertising needs and the supply of the user's webpage visits, which will boost the publisher's revenue in a long run.

\begin{figure}[t]
\centering
\includegraphics[width=1\linewidth]{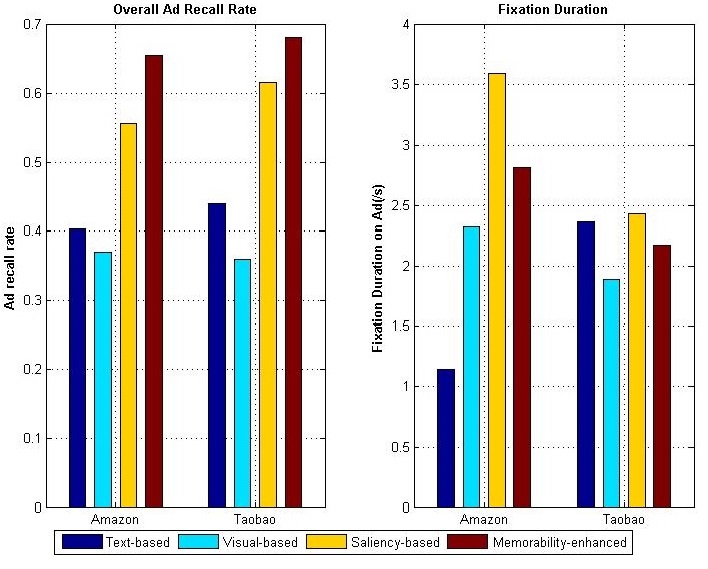}
\vspace{-15pt}
\caption{User's engagement on displayed ad under different combination of multimedia metrics. }
\label{fig:user_study}
\end{figure}

\begin{table*}[tp]
\centering
\caption{Specifications of metric variables for Stage II.}
\label{tab:second_stage_var_spec}	
\vspace{-7pt}
\begin{tabular}{l|l|l|l|l}
\hline
\multirow{2}{*}{Variable}   & \multicolumn{3}{c|}{Computation} & \multirow{2}{*}{Input for Stage II}\\
\cline{2-4}
           & Method & Input & Output         & \\
\hline
Publisher's revenue    & Stage I      & Bid & Payment & $x_{1,i,z} = $ normalized $p_{i,z}$\\
Advertiser's utility    & Stage I      & Bid, payment & Utility    & $x_{2,i,z} = $ normalized $(v_{i,z} - p_{i,z})$\\
Ad memorability & MemNet~\cite{khosla2015understanding} & Ad image & Memorability score & $x_{3,i,z} = $ normalized score\\
Ad CTR    & Given by data         & CTR &  CTR & $x_{4,i,z} = $ normalized CTR\\
Contextual relevance & TakeLab~\cite{vsaric2012takelab} & Title, keywords, description & Semantic similarity score & $x_{5,i,z} = $ normalized score\\
Ad saliency & MBS~\cite{zhang2015minimum} & Web page snapshot, ad image & Saliency map and ratio & $x_{6,i,z} = $ normalized ratio \\
\hline 
\end{tabular}
\vspace{7pt}
\end{table*}

\section{Proposed framework}
\label{sec:approach}
We propose to incorporate multimedia metrics into RTB to improve the effectiveness of display advertising. Our framework consists of two stages (see Fig.~\ref{fig:schematic_view}). The first stage is based on the second-price auction model. And the second stage is based on the optimal re-ranking model to select the most appropriate ad. Since the auction model is simple but effective, we focus on the optimal re-ranking model. When adopting the weighted linear combination model, the key issues are: what metrics should be considered and how to determine the optimal weights.  

\subsection{Select Multimedia Metrics}
\label{sec:multimedia_metrics}
As the user is the potential customer to the advertiser, any ad that can deliver the advertiser's information to the user can be viewed as an effective ad. It has been recognized that the user's acceptance of a message is sequence of psychology process: attention, comprehension, yielding, retention and action~\cite{alwitt1985psychological}. Motived by this finding, we suggest the following three multimedia metrics that can help improve the effectiveness of online advertising.

\textbf{Contextual relevance} has been widely discussed in computational multimedia advertisement~\cite{mei2010contextual}. Contextual relevance refers to the similarity between the content of the host webpage and the ad description provided by the advertiser. It has been proved to be effective in increasing the ad CTR and meanwhile maintaining user's online experience.

\textbf{Visual saliency} measures whether the ad image can be easily spotted out within the host webpage. Through a series of eye-tracking experiments, recent research found that: users intentionally avoid looking at the banner ads~\cite{sajjacholapunt2014influence}, and they tend to ignore ads located on the bottom and right area~\cite{owens2011text}. By introducing visual saliency, we can increase the probability that the user will notice the displayed ad intentionally or unintentionally.

\textbf{Image memorability} measures how likely the user will recall the ad correctly. It is an important metric for ad branding conception. It has been shown that image memorability is a stable and intrinsic property of images that is shared across different viewers. By introducing image memorability, the advertiser's benefit will be improved.

To demonstrate the effectiveness of our introduced multimedia metrics, we proposed a novel multimodal framework for contextual video advertising~\cite{xiang2015salad}. Given an online video and a set of ad candidates, we first roughly identify a set of relevant ads based on textual information matching. We then carefully select a set of candidates based on the visual content matching. In this regards, our selected ads are contextually relevant to the video in terms of both textual information and visual content. We finally select the most salient ad as the most appropriate one. We further extend the proposed framework by considering the image memorability.

In the user-study experiment, we selected 20 popular YouTube videos as our test video set, and we collected the products information from Amazon and Taobao to construct the ad dataset. We invited 80 participants to join our experiment, and we used an eye-tracker device to record their eye gaze information. Since user's behavior is sensitive to the purpose of the experiment, they did not have prior knowledge about the experiment. We also built a browser interface which is similar to YouTube to simulate the real-world online video viewing environment. The experimental result is shown in Fig.~\ref{fig:user_study}. By comparing the saliency-based advertising strategy to the text-based and visual-based strategy, we can find that: the average fixation duration on the displayed ad increases with introducing visual saliency. 
By comparing the memorability enhanced advertising to the text-based and visual-based strategy, we can find that: the average ad recall rate increases with considering image memorability. By comparing the memorability-enhanced advertising to saliency-based advertising, we observe an interesting finding: higher fixation duration does not ensure higher ad recall rate. This finding indicates that the memorability is an intrinsic property of images, which is consistent with the results in ~\cite{khosla2015understanding}.

\subsection{Optimal Re-ranking Model}
\label{sec:re-ranking}
Besides the multimedia metrics discussed in the previous section, we further consider three metric variables that is commonly discussed in online advertising: the publisher's revenue, the advertiser's utility and the ad CTR. Revenue is always the primary concern in display advertising. If advertiser is selected as the winning bidder, the revenue is her paid price, which equals to the bidding price of the next bidder. The utility shows the advertiser's cost saving. It is defined as the difference between her value and payment. The ad CTR is the more direct variable to show the effectiveness of online advertising. It can be obtained from the historical data. Table~\ref{tab:second_stage_var_spec} specifies all the metric variables as the inputs of the second stage. 

All advertisers are re-ranked in the second stage. Given a set of advertisers in auction $z$,the rank score of advertiser $i, i \in z$, is defined as follows:
\begin{align}
rs_{i, z} = & \ \sum_{k \in K} \omega_k x_{k, i, z},
\end{align}
where $x_{k, i, z}$ is the input value of metric variable $k$, $\omega_k$ is the corresponding weight. 

The optimal weights can be obtained from the historical data. The weights of variables in the rank score represent the publisher's preferences towards the metrics. As described earlier, the publisher needs to sacrifice a certain amount of revenue in the short term in order to increase the benefits of other stakeholders. Let $\theta_{1}$ be the maximum pre-determined loss rate of revenue and let $\theta_{k},\forall k \in K\backslash\{1\}$, be the minimum increase rate of other variables. Given the maximum loss and the minimum increase targets, we can obtain the optimal weights from the training set $\widetilde{Z}$. 

Our algorithm can be expressed as follows:   
\begin{align}
\max_{\omega_1, \cdots,\omega_6} 
& \ \sum_{z \in \widetilde{Z}} rs_{*,z}, \\[0.02in]
\text{subject to} 
& \ 0 \leq \omega_{k} \leq 1, \forall k \in K, \label{eq:weight_nonnegative}\\[0.02in]
& \ \sum_{k \in K} \omega_{k} = 1, \label{eq:weight_unity}\\[0.05in]
& \ |\xi_{1, \widetilde{Z}}| \leq |\theta_{1}|, \theta_1 \leq 0, \label{eq:threshold_revenue} \\[0.02in]
& \ \xi_{k, \widetilde{Z}} \geq \theta_{k}, \theta_k \geq 0, \forall k \in K\backslash\{1\}, \label{eq:threshold_others} 
\end{align}
where $rs_{*,z}$ is the rank score of the selected advertiser in auction $z$ by our proposed model, $\xi_{k, \widetilde{Z}}$ is the mean of changes of variable $k$ between our proposed model and the ground truth, defined by
\begin{equation}
\xi_{k, \widetilde{Z}} = \frac{\sum_{z\in \widetilde{Z}}(x_{k, *, z}-x_{k,\neg,z})}{\sum_{z \in \widetilde{Z}}x_{k,\neg, z}}, \label{eq:changes_of_variables}
\end{equation}
where $x_{k,*,z}$ is the input value of metric variable $k$ for the selected advertiser in auction $z$ by our proposed model, $x_{k,\neg,z}$ is the input value of metric variable $k$ for the selected advertiser in auction $z$ in the ground truth. 

The optimal weights maximize the sum of rank scores of the select advertisers from all auctions in the training set. Eqs.~(\ref{eq:weight_nonnegative})-(\ref{eq:weight_unity}) ensure each variable has an impact in the re-ranking but its impact has an upper bound. Eqs.~(\ref{eq:threshold_revenue})-(\ref{eq:threshold_others}) further specify the lower bounds of trade-offs where the maximum decrease of the publisher's revenue and the minimum increases for other variables. 

\section{Experiment}
\label{sec:experiments}
\begin{figure}[t]
%
\includegraphics[width=0.95\linewidth]{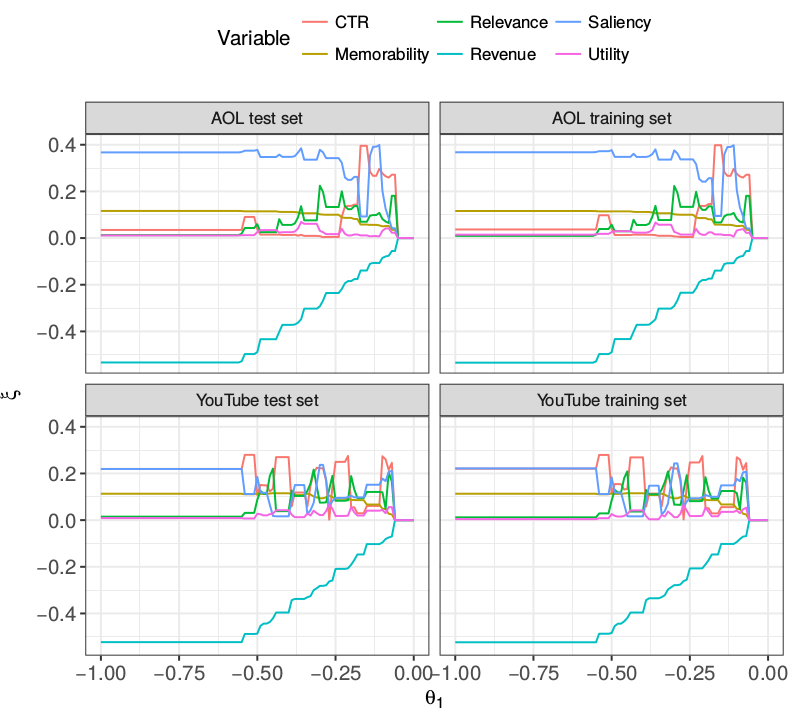}
\caption{The effect of $\theta_1$ on the changes of variables, where X-axis represents the decrease threshold of revenue ($\theta_1$) and Y-axis represents the changes of variables when compared with existing RTB system. }
\vspace{7pt}
\label{fig:theta_1_variables}
\end{figure}

To validate the proposed framework, both bidding information (e.g., bid price, ad CTR) and multimedia information (e.g., text for contextual relevance matching, ad image for visual saliency and image memorability) for each ad are needed. We use two independent datasets to simulate the real-world advertising system: the multimedia dataset (data collected from AOL and YouTube) represents the interaction between the publisher and the user, and the auction dataset shows the interplay between the publisher and the advertiser. We compare the proposed framework with the GSP auction model, which has been widely used in existing RTB system. Since revenue is always the major concern in online advertising, it will be interesting to investigate how the loss of revenue will affect the other metrics. In this experiment, we set $\theta_1$ to a non-positive value and $\theta_k=0,k=2\dots6$. We conduct 10-fold cross validation on the both AOL and YouTube dataset and the results are shown in Fig~\ref{fig:theta_1_variables}. 

From this figure, we can conclude that there exist trade-offs among stakeholders in the proposed framework, and the trade-offs change with the publisher's specific needs (the threshold values $\theta_k,k=1\dots6$). We have the following observations: 1) the changes of variables in the test set is similar to that of the training set on both AOL and YouTube multimedia datasets. The consistent results confirm our cross-validation; 2) when $\theta_1$ is close to 0, all variables remain unchanged. This is because our weight determination model (see section~\ref{sec:re-ranking}) is not able to find a solution to the constrains. In other words, the publisher is too stingy to lose revenue and too greedy to achieve improvement on the other variables. In this scenario, we will use the traditional RTB system to select the ad with highest bid price; 3) when $\theta_1$ further decreases, the revenue follows a monotone decreasing pattern and the other variables increase to various degrees. According to our optimal weights determination model, the solution space will increase with the decrease of $\theta_1$. We will select the optimal weights with the highest accumulated rank score; 4) when $\theta_1$ decreases below a threshold value, the changes of metrics remains stable. Note that, the optimal weights changes with the given threshold values. Through analyzing the historical data and the changes of variables will help the publisher decide a threshold of revenue loss.

\vspace{-4pt}
\section{Work in progress}
\label{sec:work_in_progress}
The framework described above is only applicable to the situation where each webpage has a single ad slot. We are currently working on a generalized framework where a webpage has multiple ad slots. Recently, we have constructed another multimedia dataset from Yahoo and MSN. We find that: the percentage of single slot webpage is about 18.3\% (3,390 out of 18,447) in Yahoo, and 13.8\% (1,484 out of 10,717) in MSN, respectively. The multi-slot webpage makes the computational framework more complex. In the current RTB setting, these ad slots are sold separately through individual auctions~\cite{Ben_Zwi_2015}. The following scenarios are likely to happen: 1) competitive ads are displayed on the same webpage (e.g., Apple Iphone 7 and Samsung Galaxy S8 are displayed in the same webpage about mobile techniques). The competitive ads have a negative effect on user's brand conception~\cite{burke1988competitive}; 2) an ad with a higher bid price will be placed at a less salient slot. The eye-tracking experiment has shown that user's web surfing eye-gaze follows the 'F' shape~\cite{jakob2006f}. This finding indicates the importances of different ad slot locations within the webpage. We will propose a computational framework that considers both the webpage-ad context and the ad-ad pair context. The limitations above can be acted as the hard/soft constraints.      

\section{Conclusion}
\label{sec:conclusion}
In this paper, we propose a computational framework to improve the effectiveness of display advertising by incorporating multimedia metrics. The metrics are: the contextual relevance to ensure user's online experience as well as increase the ad CTR, the visual saliency to draw user's attention and the image memorability to enhance user's brand conception. We first conduct an eye-tracking experiment to demonstrate the effectiveness of the introduced metrics. After that, we discuss how to integrate the multimedia metrics into RTB. We also propose an optimization model to determine the optimal weights of the metrics. Our experimental results show that the proposed framework is able to increase the benefits of the advertiser and the user with just a slight decrease in the publisher's revenue. However, better engagements of advertisers and users will increase the demand of advertising and supply of webpage visits, which will boost the publisher's revenue in the long run.  

\section{Acknowledgement}
This research is supported by the National Research Foundation, Prime Minister's Office, Singapore under its International Research Centre in Singapore Funding Initiative.

\newpage
\bibliographystyle{abbrv}
\bibliography{mybib}

\end{document}